%Paper: gr-qc/9407020
%From: Tevian Dray <tevian@MATH.ORST.EDU>
%Date: Fri, 15 Jul 94 16:37:43 PDT

%**start of header

% The following settings are for Plain TeX:
\magnification=\magstep1       	% Imitate 12 pt output from Plain TeX.
\font\bigbold=cmbx10 scaled 1200
\def\Bbb#1{{\cal #1}}          	% Imitate blackboard bold.

% The following settings use Tevian's twelvepoint macros:
%\input tw			% Tevian's macros for true 12 pt fonts.
%\twelvepoint			% Turn on 12 pt fonts.

\hyphenation{Berg-mann}

\vsize=9truein
\nopagenumbers			% Use page numbers at top after page 1
\headline={\ifnum\pageno=1{\hss}\else{\hss\rm -~\folio~- \hss}\fi}

\input psfig			% written with gr-qc version, namely 1.2

\newcount\EQNO      \EQNO=0
\newcount\FIGNO     \FIGNO=0
\newcount\REFNO     \REFNO=0
\newcount\SECNO     \SECNO=0
\newcount\SUBSECNO  \SUBSECNO=0
\newcount\FOOTNO    \FOOTNO=0
\newbox\FIGBOX      \setbox\FIGBOX=\vbox{}
\newbox\REFBOX      \setbox\REFBOX=\vbox{}
\newbox\RefBoxOne   \setbox\RefBoxOne=\vbox{}

%% "\normal" should be defined to restore the correct size to references which
%%           occur e.g. inside small-sized footnotes.
%%           The following line insures that a default exists.
%%           (Based on Exercise 7.7 in the TeXbook.)
\expandafter\ifx\csname normal\endcsname\relax\def\normal{\null}\fi

\def\Eqno{\global\advance\EQNO by 1 \eqno(\the\EQNO)%
    \gdef\label##1{\xdef##1{\nobreak(\the\EQNO)}}}
\def\Fig#1{\global\advance\FIGNO by 1 Figure~\the\FIGNO%
    \global\setbox\FIGBOX=\vbox{\unvcopy\FIGBOX
      \narrower\smallskip\item{\bf Figure \the\FIGNO~~}#1}}
\def\Ref#1{\global\advance\REFNO by 1 \nobreak[\the\REFNO]%
    \global\setbox\REFBOX=\vbox{\unvcopy\REFBOX\normal
      \smallskip\item{\the\REFNO .~}#1}%
    \gdef\label##1{\xdef##1{\nobreak[\the\REFNO]}}}
\def\Section#1{\SUBSECNO=0\advance\SECNO by 1
    \bigskip\leftline{\bf \the\SECNO .\ #1}\nobreak}
\def\Subsection#1{\advance\SUBSECNO by 1
    \medskip\leftline{\bf \ifcase\SUBSECNO\or
    a\or b\or c\or d\or e\or f\or g\or h\or i\or j\or k\or l\or m\or n\fi
    )\ #1}\nobreak}
\def\Footnote#1{\global\advance\FOOTNO by 1
    \footnote{\nobreak$\>\!{}^{\the\FOOTNO}\>\!$}{#1}}
\def\SameFootnote{$\>\!{}^{\the\FOOTNO}\>\!$}

\def\References{\bigskip\centerline{\bf REFERENCES}
                \smallskip\copy\REFBOX}
\def\NewRefPage{\setbox\RefBoxOne=\vbox{\unvcopy\REFBOX}%
		\setbox\REFBOX=\vbox{}%
		\def\References{\bigskip\centerline{\bf REFERENCES}
                		\nobreak\smallskip\nobreak\copy\RefBoxOne
				\vfill\eject
				\smallskip\copy\REFBOX}%
		\def\NewRefPage{}}

%%%%%%%%%%%%%%%%%%%%%%%%%%%%%%%%%%%%%%%%%%%%%%%%%
%% end Tevian's macros for automatic numbering %%
%%%%%%%%%%%%%%%%%%%%%%%%%%%%%%%%%%%%%%%%%%%%%%%%%

%\input number4			% Stuart's additions (TD's version)
%% FOR CHAPTERS AND SECTIONS.  NUMBERING BY CHAPTERNO.ITEMNO

%%%% modified 6/15/94 by TD to be compatible with number.tex

%%%% modified 6/6/93 by SB for use of itemno
%%%% i.e. number equations, theorem, lemma, etc by section#.item#

%\input number

\def\Fig#1{\global\advance\FIGNO by 1 Figure~\the\FIGNO: {#1}%
    \global\setbox\FIGBOX=\vbox{\unvcopy\FIGBOX
      \narrower\smallskip\item{\bf Figure \the\FIGNO~~}#1}}

\newcount\ITEMNO     \ITEMNO=0

%\def\itemno{\global\advance\ITEMNO
%by 1 \the\SECNO .\the\ITEMNO}
%by 1 \the\ITEMNO.}
\def\itemno{\global\advance\ITEMNO by 1 \the\ITEMNO}

%%%%%%%%%%%%%%%%%%%%%%%%%%%%%%%%%%%%%%%%%%%%%%%%%
%%%% Theorem, Proposition, Definition, Claim
%%%% Formats
%%%%%%%%%%%%%%%%%%%%%%%%%%%%%%%%%%%%%%%%%%%%%%%%%

% Stuart's macros

%\input /users/boersma/thesis/inputs/mymacros

%Misc
\def\cross{\times}

%Time derivatives

%Many different Christoffel symbols of the second kind
\def\Gamu#1#2#3{\Gamma^{#1}_{\ #2#3}}

\def\Gamperpu#1#2#3{{^\perp}\!\Gamma^{#1}_{\ #2#3}}

%\def\SGamu#1#2#3{{\cal G}^{#1}_{\ #2#3}}

%Fractions and (first and second) partial derivatives
\def\frac#1#2{{{#1}\over {#2}}}

\def\at#1{\lower.8ex\hbox{${\big\vert _{#1}}$}}

%more derivatives

\def\starry#1{{_{*#1}}}
\def\pel#1{{\rm P}_i}
\def\dt{\frac{\partial}{\partial t}}

\def\dxi{\frac{\partial}{\partial x^i}}

\def\dxj{\frac{\partial}{\partial x^j}}

\def\dxalpha{\frac{\partial}{\partial x^\alpha}}
\def\dxbeta{\frac{\partial}{\partial x^\beta}}
\def\dyalpha{\frac{\partial}{\partial y^\alpha}}
\def\dybeta{\frac{\partial}{\partial y^\beta}}
\def\partiali{\partial_i}

\def\partialt{\partial_t}

	%Lie differentiation
\def\lstar#1{\hbox{\it
\$}_{\!\!*\,\,\lower2pt\hbox{$\!\!\scriptscriptstyle#1$}}}

%Covariant Derivatives and Nablas

\def\sdel#1{{\nabla_{\!\!*#1}}}

\def\d#1{D_{#1}}

%some tensors

\def\rperpu#1#2#3#4{{^\perp\! R}^{#1}_{\ #2#3#4}}

\def\zelmanovu#1#2#3#4{Z^{#1}_{\ #2#3#4}}

\def\D{{\cal D}}				%%deficiency

\def\tordperp{^\perp\!T_{\!D}}

\def\proj#1#2{P_{#1}^{\ #2}}

%Laplacians
\def\sqr#1#2{{\vbox{\hrule height.#2pt
              \hbox{\vrule width.#2pt height#1pt \kern#1pt
                   \vrule width.#2pt}\hrule height.#2pt}}}

%metrics anyone?

\def\hij{h_{ij}}
\def\gij{g_{ij}}
\def\kij{k_{ij}}

\def\bracket#1#2{\left[ #1 , #2 \right]}

%parametric tensors
\def\p1tensors{{\cal P}\! T_1}

%parametric actions

%misc

\def\vfs{\raise3pt\hbox{$\chi$}(\M)}
\def\pvfs{\raise3pt\hbox{$\chi_{_*}$}(\Sigma)}
		%set of functions
	%set of parametric functions

%maps

\def\push{{_{*}}}
\def\piinverse{\pi^{-1}}

%jets

\def\j1pi{J^1_{\pi}}

  %sections of a bundle
\def\hforms{\bigwedge\,\raise6pt\hbox{$\!\!\scriptstyle
1$}\lower6pt\hbox{$\!\!\scriptstyle 0$} \pi}

%manifolds
\def\M{{\cal M}} 		%a spacetime

 		%leaves of a foliation
		%foliation

\def\picture#1#2#3{\psfig{figure=#1,height=#2,width=#3}}

\def\BR{{\Bbb R}}

%**end of header

\def\today{\number\day\space\ifcase\month\or
  January\or February\or March\or April\or May\or June\or
  July\or August\or September\or October\or November\or December\fi
  \space\number\year}
%\rightline{gr-qc/9407012}
\rightline{15 July 1994}
%\rightline{DRAFT,\space\today}
\bigskip\bigskip

\null\bigskip\bigskip\bigskip
\centerline{\bigbold SLICING, THREADING \& PARAMETRIC MANIFOLDS}
\bigskip

\centerline{Stuart Boersma}
\centerline{\it Department of Mathematics, Oregon State University,
		Corvallis, OR  97331, USA
\Footnote{Present address: Division of Mathematics and Computer Science,
Alfred University, Alfred, NY  14802}
}
\centerline{\tt boersma@math.orst.edu}
\medskip
\centerline{Tevian Dray}
\centerline{\it Department of Mathematics, Oregon State University,
		Corvallis, OR  97331, USA}
\centerline{\tt tevian@math.orst.edu}

\bigskip\bigskip\bigskip\bigskip
\centerline{\bf ABSTRACT}
\midinsert

\narrower\narrower\noindent
We present a unified treatment of the {\it slicing} (3+1) and
{\it threading} (1+3) decompositions of spacetime in terms of foliations.
It is well-known how to decompose the metric and connection in
the slicing picture; this is at the heart of any initial-value problem in
general relativity.  We describe here the analogous problem in the threading
picture, recovering the recent results of Perj\'es on {\it parametric
manifolds}.

\endinsert
\vfill
\eject

\Section{Introduction}

The main purpose of this paper is to present a unified treatment of the
well-known {\it slicing} or ADM formalism with its less well-known dual
formalism, which we call {\it threading}.  In the slicing viewpoint, spacetime
is foliated with spacelike hypersurfaces; in the threading viewpoint spacetime
is foliated with timelike curves.  Slicing and threading are equivalent in the
special case where both foliations exist and are orthogonal to each other; the
interesting case is when the curves are not hypersurface-orthogonal.  The
slicing viewpoint corresponds to a global time-synchronization, whereas the
threading viewpoint treats a family of observers as fundamental.  We show here
how both slicing and threading can be obtained from a more general
decomposition of the metric.

In the slicing viewpoint, one can regard tensor fields as time-dependent
tensors on a particular hypersurface.  One can then study the intrinsic and
extrinsic geometry of the surface, for instance when formulating an initial
value problem.  This is not as simple from the threading viewpoint due to the
likely absence of any hypersurfaces orthogonal to the given curves.  One is
thus naturally led to {\it parametric manifolds}, in which one considers
time-dependent tensors on the manifold of orbits of the threading curves
\Ref{{Z. Perj\'es}, {\it The Parametric Manifold Picture of Space-Time},
{Nuclear Physics} {\bf B403}, 809 (1993)}\label\PERJES
.  This is equivalent to using a projected geometric structure on a
hypersurface of constant time, rather than the intrinsic geometry such a
surface inherits from spacetime.
\Footnote{As discussed in more detail below, the threading metric corresponds
to the distance between nearby observers, as measured orthogonally to the
observers.  This is in general different from the distance as measured on a
hypersurface of constant time.} This parametric viewpoint is closely related
to Kaluza-Klein theories.  Both can be viewed as projections of a higher
dimensional geometry into a lower dimension.  This perspective allows us to
give the theory of parametric manifolds a solid mathematical foundation.
Central to this discussion is a new, torsion-like quantity, the {\it
deficiency}, which measures whether the given family of observers is
hypersurface orthogonal.

We begin with a review of slicing in Section 2.  Although this
material is well-known, we give a detailed presentation of two different
interpretations of slicing in order to set the stage for the completely
analogous presentation of threading in Section 3.  In Section 4 we discuss the
general problem of decomposing the metric on a fibre bundle into metrics on
the base space and on the fibres, and we then show how to recover both slicing
and threading as special cases.  An overview of parametric manifolds is given
in Section 5, and a discussion of possible applications appears in Section 6.

%%%%%%%%%%%%%%%%%%%%%%%%%%%%%%%%%%%%%%%%%%%%%%%%%%%%%%%%%%%%%%%%%%%%%%%%%%%%%%

\Section{Slicing}

The (3+1)-decomposition of a 4-dimensional spacetime is the standard
framework for formulating the dynamics of geometry ({\it cf.}\
\Ref{{C. Misner, K. Thorne, J. Wheeler}, {\bf Gravitation}, {W. H. Freeman and
Company, San Fransisco, 1973}}\label\MTW
).
There exist two standard approaches to such a splitting, both of which yield
the standard definitions of {\it lapse} and {\it shift}, one being a
construction process and the other a decomposition process.  For the
construction, one begins with 3-dimensional surfaces and attempts to ``fill
in'' between these surfaces to construct a spacetime which admits the original
3-dimensional surfaces as a foliation of spacelike hypersurfaces.  The
spacetime metric is thus constructed out of the 3-metric of the hypersurfaces
as well as additional bits of information.  Alternatively, one could start
with a spacetime which admits a 1-parameter family of spacelike hypersurfaces
and then decompose all of the original 4-dimensional geometrical information
({\it e.g.}\ tensor fields) into two pieces, one tangent to the surfaces and
one normal to the surfaces.  As we shall see, both approaches yield an
equivalent (3+1)-interpretation of spacetime.

Throughout the next two sections we will be working with complete spacetimes
foliated by spacelike hypersurfaces.  We will further assume that the
hypersurfaces are all diffeomorphic to each other.  Thus, one may simplify the
notation by working in extremely nice coordinate neighborhoods.  Begin by
introducing a global time function $t$ which can be regarded as the parameter
which labels the hypersurfaces.  We will work in a neighborhood small enough
so that the intersection of each hypersurface with the neighborhood admits
coordinates $\{ x^i\} =\{ x^1,\ x^2,\ x^3\}$ on the hypersurface $\Sigma_t$.
Thus, $p\in\Sigma_t$ can be given the coordinates $(t,x^i).$ To simplify the
notation, we will use Greek letters as indices which take on all four
spacetime dimensions with $x^0 \equiv t$.  Thus, $\{x^\alpha\}=\{t,x^i\}$ and
we can write the spacetime metric $g$ in terms of its components given by
$ds^2=g_{\alpha\beta}\, dx^\alpha dx^\beta$.

\Subsection{The Construction}

Suppose one had a spacetime foliated by a 1-parameter family of spacelike
hypersurfaces, $\Sigma_t$.  One would like to realize the 4-geometry of this
spacetime as arising from the 3-geometries of these surfaces.  Thus, one would
like to ``construct'' the spacetime metric out of the spatial metrics of the
surfaces.  Of course, additional information must also be provided.  Following
the description in \MTW , let us assume that the 3-geometry of two
infinitesimally close surfaces is known.  Label these surfaces by $\Sigma_t$
and $\Sigma_{t+\Delta t}$.  Each of these surfaces has an associated spatial
metric, $k_t$ and $k_{t + \Delta t}$.  At the risk of de-emphasizing the
dependence on the coordinate $t$, we will use the same notation to refer to
both spatial metrics and write $k_{ij}\, dx^i dx^j$ for the 3-metrics on both
surfaces.

We now describe the 4-geometry that fills in between these slices.  Given
a point $p_0
=(x^i,t)\in\Sigma_t$ and a nearby point $q_0=(x^i+\Delta x^i , t+\Delta
t)\in\Sigma_{t+\Delta t}$, we are interested in calculating the coordinate
distance $d(p_0,q_0)$ between $p_0$ and $q_0$.  By taking advantage of the
existence of a metric in the slice $\Sigma_t$, it seems most natural to use
the Pythagorean Theorem of Lorentzian geometry and write
$$d(p_0,q_0)^2 = d(p_0,p_1)^2 - d(p_1,q_0)^2 \Eqno$$\label\PythagThm
where $p_1 \in \Sigma_t$ is chosen so that $d(p_1,q_0)$ is the orthogonal
distance between the two hypersurfaces.  See Figure 1.

%\vskip.25in
\bigskip
{\offinterlineskip
\centerline{\picture{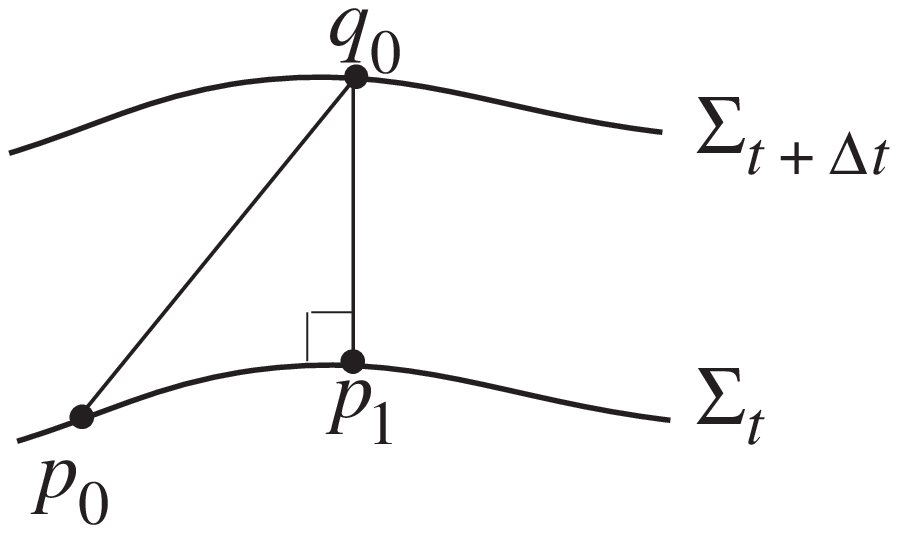}{1.5in}{2in}}
%\vskip-12truept
\centerline{\Fig{Calculating Distances}}
}
\bigskip
%\vskip.25in

It is now apparent that in order to fully construct a spacetime metric much
information must still be specified.  As we have no {\it a priori} knowledge
of what it means to move orthogonally to the surfaces, the location of $p_1$
is not fully determined.  As $\Delta t$ approaches zero, the point $q_0$
should approach $p_1$.  However, there is no reason to assume that the
coordinates of $p_1$ are $(x^i + \Delta x^i , t)$.  Rather, the point $p_1$
could be ``shifted'' in any of the three spatial directions.  Thus, we assign
the coordinates $p_1 = (x^i + \Delta x^i + N^i\Delta t , t)$.  The three
functions $N^i$ depend on the coordinates of $\Sigma_t$ as well as the
parameter $t$.  As the functions $N^i$ describe how the nearby surfaces are
shifted with respect to one another, they are commonly referred to as the
components of the {\it shift vector}.  The given metric $\kij$ may now be used
to measure $d(p_0,p_1)$.

The quantity $d(p_1,q_0)$ is still not determined.  In order to fix this
distance, one must know the relationship between the proper time from
$\Sigma_t$ to $\Sigma_{t+\Delta t}$ and the arbitrary parameter $t$.  Again,
this distance may depend on the coordinates in $\Sigma_t$ as well as $t$.
Define the {\it lapse function} $N$ by
$$d(p_1,q_0) = N(x^i,t)\, \Delta t.$$

One may now describe the 4-geometry in terms of the lapse function and shift
vector.  Adding these newly defined quantities to Figure 1 yields the picture
shown in Figure 2.

%\vskip.25in
\bigskip
{\offinterlineskip
\centerline{\picture{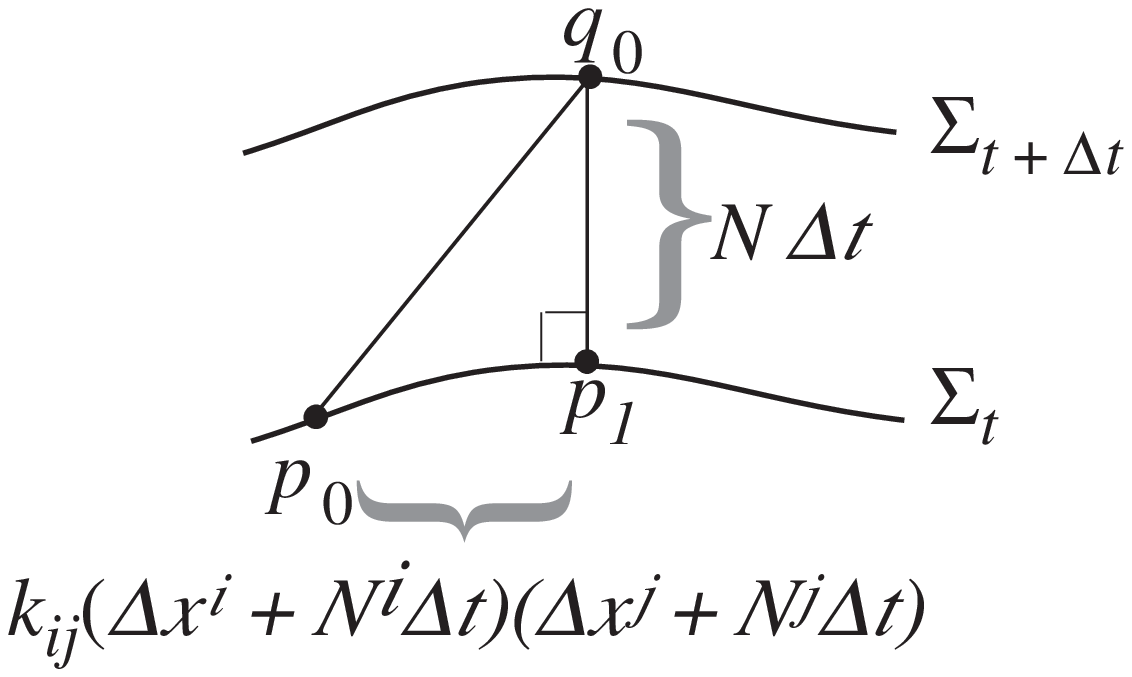}{1.75in}{2.5in}}
%\vskip-12truept
\centerline{\Fig{Slicing Lapse Function and Shift Vector}}
}
\bigskip
%\vskip.25in

Using equation \PythagThm\ and letting $\Delta t \rightarrow 0$ we see that
the 4-geometry of
the spacetime can be represented by the line element
$$\eqalign{ds^2 &= k_{ij}(dx^i+N^idt)(dx^j+N^jdt) - N^2 dt\cr
                &= (N_iN^i - N^2)\,dt^2 + 2N_i\,dt\,dx^i +
                k_{ij}\,dx^i\,dx^j\cr}\Eqno$$\label\slelt
where we have used the 3-metric $k_{ij}$ to define $N_i=k_{ij}N^j.$  As
mentioned earlier, the functions $N^i$ are thought of as the component
functions of a vector field ``tangent'' to each hypersurface.  The {\it shift
vector field} is defined by
$$N^i\dxi.$$
Thus each hypersurface has a spatial metric, $\kij$, a tangent vector
field, $N^i \dxi$, and a function $N$.  As we have seen, these three spatial
quantities may be used to construct the 4-dimensional spacetime metric $g$.

In matrix notation, one can write the components of the spacetime metric
tensor in
terms of $N,\ N^i$, and $k_{ij}$ as follows:
$$(g_{\alpha\beta}) = \left(\matrix{-(N^2-N_mN^m)&N_j\cr
			\noalign{\hbox{\strut}}
			N_i&k_{ij}\cr}\right)$$
with inverse
$$(g^{\alpha\beta}) = \left(\matrix{-N^{-2}&N^{-2}N^j\cr
			\noalign{\hbox{\strut}}
			N^{-2}N^i&k^{ij}-N^{-2}N^iN^j\cr}\right)$$
where $k^{ij}$ is the inverse of $k_{ij}$ defined by
$$k_{il}k^{lj}=\delta_i^{\ j}.$$

Our parameter $t$ now takes the role of a spacetime coordinate whose
coordinate vector field $\dt$ may be interpreted as representing the ``flow of
time'' in the newly constructed spacetime.
Since the coordinates $x^i$ are constant along integral curves of $\dt$, one
may think of the lapse and shift as the means of identifying points on
different hypersurfaces.  See Figure~3.

\bigskip
{\offinterlineskip
\centerline{\picture{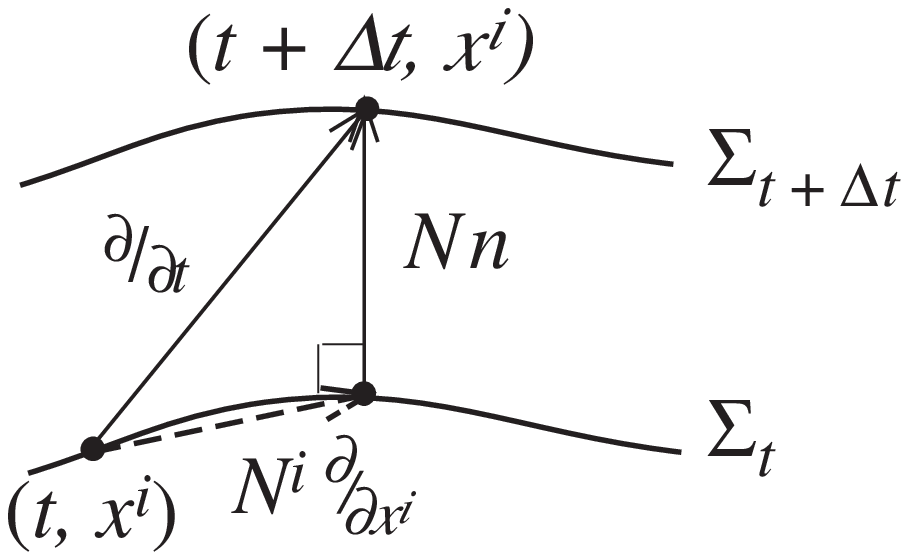}{1.75in}{2.5in}}
%\vskip-12truept
\centerline{\Fig{Decomposition of $\dt$}}\label\SlFigC
}
\bigskip

The shift vector field $N^i\dxi$ was defined to account for the fact that
there was no {\it a priori} knowledge of directions orthogonal to the surfaces
$\Sigma_t$.  However, in terms of the shift vector and lapse function we may
now easily describe a future pointing unit spacetime vector field normal to
each surface.  Call this normal vector field $n$.
Using $\langle\ ,\ \rangle$ to represent the 4-metric we just constructed,
we observe
that $\left<Ndt,\ Ndt\right>=-1.$  Therefore, $n$ is given by the
metric-dual of
the 1-form $N\,dt$.  Explicitly,
$$ n = \frac 1N\dt - \frac 1N N^i\dxi.\Eqno$$\label\constr
It is now apparant how the functions $N^i$ describe
the ``shifting'' of neighboring surfaces.  The case of vanishing shift
corresponds to the scenario where coordinate time is flowing orthogonally to
the
hypersurfaces ({\it i.e.}\ the directions of $n$ and $\dt$ agree).

\Subsection{The Decomposition}
As stated earlier, the above construction is simply an orthogonal splitting of
spacetime geared towards an initial value formulation of spacetime.  To see
the obvious, begin as above with a spacetime which admits a foliation of
spacelike hypersurfaces.  If we let $n^\alpha$ be the components of the
future pointing unit vector field $n$ normal to the hypersurfaces
$\Sigma_t$, the naturally induced metric on each hypersurface may be obtained
{}from the projection tensor
$$k_{\alpha\beta}= g_{\alpha\beta}+n_\alpha n_\beta \Eqno$$\label\slmetric
 where
$n_\alpha=g_{\alpha\beta}n^\beta$ ({\it cf.}\
\Ref{{R. Wald}, {\bf General Relativity}, {The University of Chicago Press,
Chicago, 1984}}\label\WALD
)  For vector fields
$X=X^\alpha\dxalpha$ and $Y=Y^\beta\dxbeta$ tangent to $\Sigma_t$,
$$\eqalign{k_{\alpha\beta}X^\alpha Y^\beta&=g_{\alpha\beta}X^\alpha Y^\beta\cr
	&=g_{ij}X^i Y^j.\cr}$$
the functions $\kij = \gij$ may thus be thought of as the components of
a 3-dimensional metric on each hypersurface.  One must not lose sight of
the fact that the functions $\kij$ depend on the spacetime coordinate $t$ (as
do the $g_{\alpha\beta}$).  We will refer to the
functions $k_{ij}$ as the components of the {\it slicing metric}.

The slicing metric on $\Sigma_t$ is the
naturally induced metric in the following sense: for each imbedding
$\iota_t:\Sigma_t\hookrightarrow \M$, $k={\iota_t}^*(g)$, where ${\iota_t}^*$
refers to the natural map on the cotangent spaces $T^*\M$ and $T^* \Sigma_t$.
Thus, $k$ is both the projection of $g$ to $\Sigma$ via \slmetric\ and the
pullback of $g$ to $\Sigma$.  Using ${\iota_t}_*$ to denote the natural map
between the tangent spaces, we
may work out the relationship explicitly.  For tangent vector
fields $X$ and $Y$ on $\Sigma_t$ we can write $X=X^i\dxi$, $Y=Y^i\dxi$, and
$$\eqalign{k_{ij} X^iY^j&={\iota_t}^*(g)_{ij} X^iY^j\cr
                        &=g_{\alpha\beta}\left({\iota_t}_*(X)\right)^\alpha
\left({\iota_t}_*(Y)\right)^\beta\cr
                        &=g_{ij}X^i Y^j .\cr}$$
One may decompose the coordinate vector field $\dt$ into vector
fields normal and tangent to each surface $\Sigma_t$. Thus one has $$\dt =
N n + N^i\dxi . \Eqno$$\label\decomp

Equation \decomp\  defines the {\it slicing lapse function} $N$ and the {\it
slicing shift vector field} $N^i\dxi$ and can be seen to agree with the earlier
definitions by comparing equation \decomp\  with \constr .  In this scenario,
the shift vector measures the tilting of $\dt$ away from the direction normal
to the hypersurfaces.

Since $N^i\dxi$ is tangent to each hypersurface, we use the slicing metric to
define $N_i = k_{ij}N^j.$

%%%%%%%%%%%%%%%%%%%%%%%%%%%%%%%%%%%%%%%%%%%%%%%%%%%%%%%%%%%%%%%%%%%%%%%%%%%%%%

\Section {Threading}

In the (3+1)-decomposition (or {\it slicing}) of spacetime, one has a
foliation of spacetime by spacelike hypersurfaces labeled by a global time
function $t$.  This time function together with the earlier definitions of the
lapse function and shift vector gives one a way of identifying points on
different hypersurfaces.  In effect, one has, in addition to a foliation of
spacetime by hypersurfaces, a congruence of curves given by the integral
curves of the coordinate vector field $\dt$.  While the spacelike nature of
the hypersurfaces is an integral part of the standard (3+1)-decomposition,
there are no similar causality conditions on the congruence of curves.
Although we usually think of the parameter $t$ as a local time coordinate, no
formal causality restriction is necessary.  When one adopts the dual ansatz of
a foliation of spacetime by timelike curves together with a foliation of
hypersurfaces (with no causality conditions imposed upon them), one is led to
consider a (1+3)-decomposition (or {\it threading}) of spacetime (see
\Ref{{R. Jantzen and P. Carini}, {\it Understanding Spacetime Splittings
and Their Relationships}, in {\bf Classical Mechanics and Relativity:
Relationship and Consistency}, ed.\ by G. Ferrarese, Bibliopolis, Naples,
185--241, 1991}\label\JANTZEN
).

In such a setting, the timelike congruence may be interpreted as the
world-lines of a family of observers, while the hypersurfaces play the
fundamental role of synchronizing the clocks of the different observers.

As in the last section, we will present the threading point of view from two
different perspectives.  First, we will address the issue of constructing a
spacetime from a given family of curves.  Second, we will illustrate the
threading point of view by considering a certain decomposition of spacetime.
One should notice the similarities between the slicing and threading points of
view.

\Subsection{The Construction}

In the previous section we saw how one would construct a spacetime metric from
a 1-parameter family of 3-dimensional Riemannian manifolds.  The resulting
4-metric was easily described in terms of the given metrics on the surfaces,
the slicing lapse function, and the slicing shift vector field.  Suppose one
is instead given a family of timelike curves.  How would one go about
constructing a spacetime which realized the original family of curves as a
congruence of timelike curves?  We will proceed as we did in the case of
slicing.

In the earlier (3+1)-construction we had a parameter which labeled each
hypersurface.  Let us assume we have parameters $x^i,i=1,2,3$ which label each
curve $L_{x^i}$.  On each curve suppose we have a coordinate $t$ as well as a
metric $l$, which can thus be expressed as
$$l=-M^2\, dt^2.$$

We will interpret these curves as being world-lines of observers, and hence
require that they be timelike.  Consider the same measuring problem as before,
that is, letting $p_0 = (x^i,t)\in L_{x^i}$ and
$q_0=(x^i + \Delta x^i, t+\Delta t)\in L_{x^i + \Delta x^i}$
we are interested in measuring the coordinate distance $d(p_0,q_0)$  between
$p_0$ and $q_0$.  Since we are assuming we can measure distances in each curve
$L_{x^i}$, again use the Pythagorean Theorem to write
$$d(p_0,q_0)^2 = -d(p_0,q_1)^2 + d(q_1,q_0)^2.$$
Here $d(q_1 , q_0 )$ is meant to refer to the orthogonal distance between two
nearby curves.  See Figure 4.

%\vskip.25in
\bigskip
{\offinterlineskip
\centerline{\picture{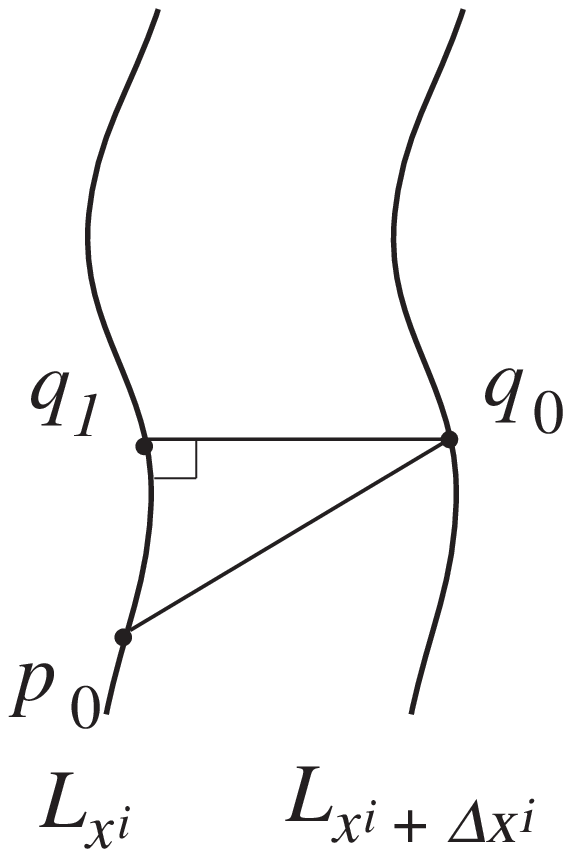}{2in}{1.5in}}
%\vskip-12truept
\centerline{\Fig{Calculating Distances}}
}
\bigskip
%\vskip.25in

Since we do not have any notion of traveling ``orthogonally'' to the curves
$L_{x^i}$, the $t$-coordinate of $q_1$ is not determined.  The position of
$q_1$ along $L_{x^i}$ is affected by each of the $\Delta x^i$.  Assign
coordinates to $q_1$ by $q_1 = (x^i,t+\Delta t - M_i \Delta x^i)$.  Again, the
$M_i$ record the amount of ``shifting'' of $q_1$ with respect to nearby
curves.  That is, the $M_i$ may be thought of recording how $L_{x^i + \Delta
x^i}$ has been shifted with respect to $L_{x^i}$ in the construction process.
Therefore, we have
$$d(p_0,q_1) = M(\Delta t - M_i \Delta x^i).$$
The three functions $M_i$ and the function $M$ depend on the parameters $x^i$
as well as the coordinate $t$.

We now need to specify the relationship between the parameters $x^i$ and the
proper coordinate distance between neighboring curves.  We thus introduce a
``spatial metric'' of the form $\hij \Delta x^i \Delta x^j$ which gives the
distance between $L_{x^i}$ and $L_{x^i + \Delta x^i}$ for various choices of
$\Delta x^i$.  While we assume that $\hij = h_{ji}$, the functions $\hij$ may
otherwise be chosen arbitrarily.  We continue our
construction of the 4-metric by assuming that this distance is precisely
$d(q_1 , q_0)$ ({\it i.e.}\ measured orthogonally).
See Figure 5.

%\vskip.25in
\bigskip
{\offinterlineskip
\centerline{\picture{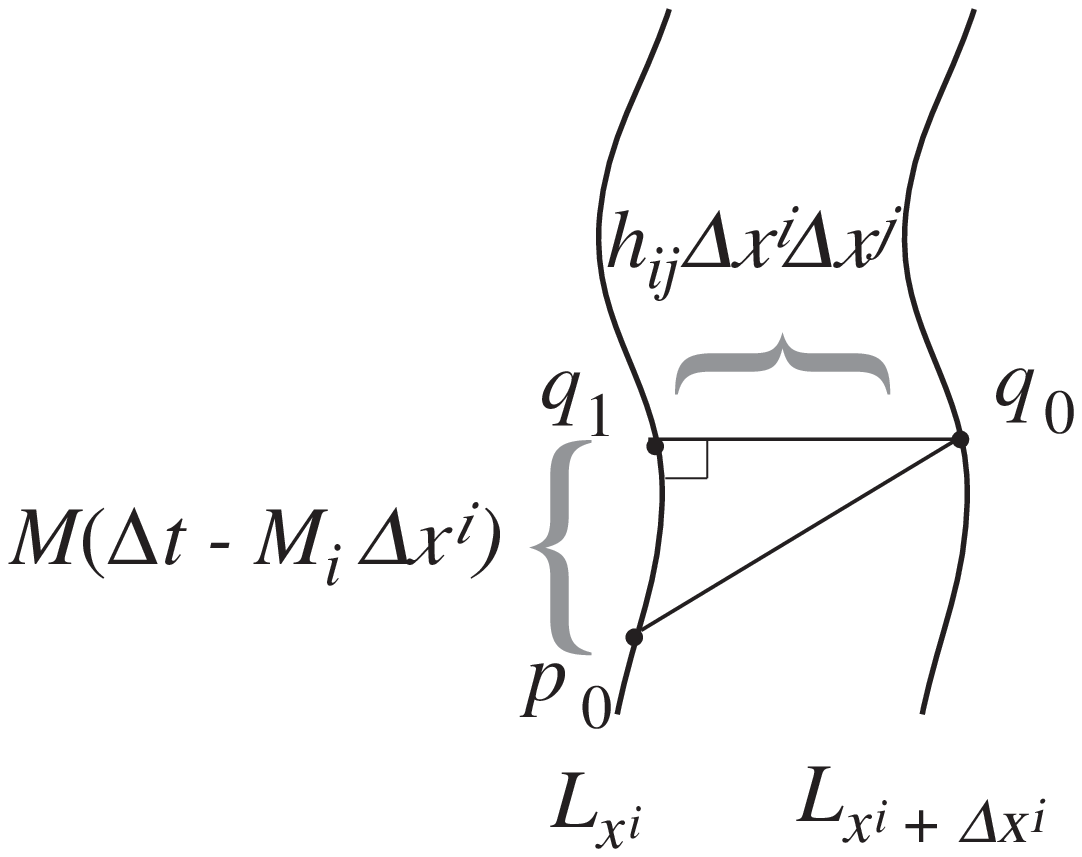}{1.75in}{2.25in}}
%\vskip-12truept
\centerline{\Fig{Threading Lapse Function and Shift 1-Form}}
}
\bigskip

%\vskip.25in
Thus, the Pythagorean Theorem implies that the spacetime metric may be written
as
$$\eqalign{ds^2 &= -M^2(dt - M_i\, dx^i)^2 + h_{ij}\, dx^i \, dx^j \cr
	&= -M^2\, dt^2 + 2M^2M_i\, dx^i \, dt
		+(h_{ij} - M^2M_iM_j)\,dx^i\, dx^j .\cr}\Eqno$$\label\ThLineElt
The component $M$ of the original metric along each curve is referred to as
the {\it threading lapse function}.  As we see in the above representation of
the metric (equation \ThLineElt ), the functions $M_i$ are most naturally
associated with the 1-form $M_i dx^i$.  The three functions $M_i$ are referred
to as the components of the {\it threading shift 1-form} $M_i\, dx^i$.
Finally, the functions $\hij$ may be thought of as the components of a metric,
the {\it threading metric}.

In the case of slicing, one thought of the slicing shift vector as a three
dimensional spatial vector field on the surfaces $\Sigma_t$ and, hence, raised
and lowered its indices with the slicing metric.  In the present case of
threading, one may again adopt the convention that the threading shift
1-form be treated as a three dimensional tensor.  Under such a convention,
the threading metric $\hij$ may be used to raise and lower its indices.

The threading shift 1-form was defined in order to introduce some notion of
traveling
``orthogonally'' to the curves $L_{x^i}$.  The unit 1-form
which annihilates the space of vectors orthogonal to the threading vector field
may be written
$$m = -M(dt - M_i \, dx^i).$$
One should compare this equation with the analogous equation for slicing
(equation \constr ).

\Subsection{\bf The Decomposition}

The threading lapse function and shift 1-form field may also be viewed as
arising from a orthogonal decomposition of spacetime.  Analogous to the
(3+1)-decomposition, the so-called (1+3)-decomposition attempts to
decompose spacetime quantities into pieces orthogonal to the given
congruence of curves, and pieces tangent to the congruence.
As above, we will work in coordinates $(t,x^i)$ where $t$ acts as a
parameter along the integral curves of $\dt$ (the {\it threading curves})
and $x^i$ are coordinates on each hypersurface $\Sigma_{t_0}=\{ t\equiv
t_0\}$.  As in \JANTZEN , we will refer to $\dt$ as the
{\it threading vector field}.

The normalization of the threading vector field is used to define the
{\it threading lapse function} $M$:
$$\left<\dt ,\dt\right> = -M^2,$$
where $\langle\ ,\ \rangle$ refers to the spacetime metric $g$.
If one views the threading curves as the world-lines of a family of
observers, the threading lapse function measures the rate of change of the
observed proper time with respect to the coordinate time function $t$.

In the slicing point of view one had to describe the discrepancy between $\dt$
and the direction normal to the hypersurfaces.  Analogously, in the present
scenario one wishes to measure the amount of tilting of the local rest spaces
of the observers with respect to each coordinate direction $\dxi$, keeping in
mind that the local rest spaces of the observers need not constitute a
hypersurface!  Following the flavor of equation \decomp , we decompose the
coordinate 1-form $dt$ into a piece which annihilates the local rest spaces
and a piece which is in the cotangent space of each surface.  Letting $m$
represent the metric dual of the unit vector field tangent to the threading
curves, we have $m(\dt)=-M$, so that $dt$ must have the form
$$dt=-\frac 1M m + M_idx^i.\Eqno$$\label\ThDecomp
The functions $M_i$ thus defined are the components of the {\it threading
shift 1-form}.

Using the above definitions of $M$ and $M_i$, the (1+3)-decomposition
of the components of the 4-metric takes the following form:
$$(g_{\alpha\beta}) = \left(\matrix{-M^2&M^2M_j\cr
\noalign{\hbox{\strut}}
M^2M_i &
h_{ij}-M^2M_iM_j\cr}\right)\Eqno$$\label\threadmet
with inverse
$$(g^{\alpha\beta})= \left(\matrix{-(M^{-2}-M_mM^m)& M^j\cr
\noalign{\hbox{\strut}}
	M^i & h^{ij}\cr}\right)$$
where we have defined $M^i=h^{ij}M_j$ and where the functions $h_{ij}$ defined
by equation \threadmet\ are the components of the {\it threading metric}, with
inverse $h^{ij}$.

Although one may take equation \threadmet\ as the definition of $h_{ij}$,
historically it has a more familiar definition.  For instance, in both
\Ref{{L. Landau and E. Lifshitz}, {\bf The Classical Theory of Fields},
{Addison-Wesley, Cambridge, 1951}}
and
\Ref{{C. M\o ller}, {\bf The Theory of Relativity}, Second Edition,
{Clarendon Press, Oxford, 1972}} \label\MOLLER
one is given a physical
interpretation of the threading metric.  In general relativity, to
calculate the spatial distance between an observer and an
infinitesimally close event, one may direct a light signal from the observer to
the event and back and calculate the ``time" of propagation.  One finds the
spatial
distance $dl$ to be given by
$$dl^2=\gamma_{\alpha\beta}dx^\alpha dx^\beta$$
where
$$\gamma_{\alpha\beta}=g_{\alpha\beta}-\frac{g_{0\alpha}g_{0\beta}}{g_{00}}.$$
Note that $\gamma_{00}=\gamma_{0\alpha}=\gamma_{\alpha 0}\equiv 0$ and
that $\gamma_{ij}\equiv h_{ij}$ ($i,j=1,2,3$) as defined earlier (in our
adapted coordinate system).
\Footnote{In
\Ref{{A. Einstein and P. Bergmann}, {Ann.~of Math} {\bf 39}, 683--701, (1938).
\hfill\break
{\it This article was reprinted in}
{\bf Introduction to Modern Kaluza-Klein Theories}, edited by T. Applequist,
A. Chodos, and P.G.O. Freund, Addison-Welsey, Menlo Park, 1987.}\label\EIN
, Einstein and Bergmann used a
similar argument during their attempts to generalize Kaluza's theory of
electricity.  Einstein and Bergmann, however, were working in a
5-dimensional space, so that in their formalism the 4-dimensional
spacetime metric took the role of the threading metric in the above
discussion.}

One can see that the threading metric simply measures what Cattaneo
\Ref{{C. Cattaneo}, {Nuovo Cimento}, {\bf 10}, {318}, (1958)} \label\CATI
refers to as the {\it space norm} of any 4-vector.  That is, $h_{ij}$
measures the norm of the component perpendicular to the threading
curves.  Specifically, for any 4-vector $V^\alpha$ write $V^\alpha =
{V_\parallel}^\alpha + {V_\perp}^\alpha$ where ${V_\parallel}^\alpha$ is
parallel to the threading curves
and ${V_\perp}^\alpha$ is perpendicular.  Letting $m^\alpha$ be the unit
vector tangent to $\dt$ one has
$$\eqalign{m_\alpha&=g_{\alpha\beta}m^\beta\cr
                        &=\frac 1{\sqrt{-g_{00}}} g_{0\alpha}\cr}.$$
Thus,
$$\eqalign{g_{\alpha\beta} {V_\perp}^\alpha {V_\perp}^\beta &=
(g_{\alpha\beta}+ m_\alpha m_\beta) V^\alpha V^\beta\cr
         &=(g_{\alpha\beta} - \frac{g_{0\alpha} g_{0\beta}}{g_{00}})
         V^\alpha V^\beta\cr
         &=(g_{ij} + M^2M_iM_j)V^i V^j\cr
         &=h_{ij} V^i V^j.\cr}$$

At this point one notices a crucial difference between the slicing and
threading pictures of spacetime.  When slicing spacetime with spacelike
hypersurfaces, one defines the slicing metric which naturally lives on these
hypersurfaces.  While the threading metric arises in an analogous way, there
exists no corresponding space (hypersurface) on which it naturally exists
(since the local rest spaces of the observers may not be surface forming).
One therefore constructs an abstract 3-manifold with the threading metric as
its Riemannian metric.  By identifying each threading curve with the point
$(t_0,x^i)$ at which it pierces the slice $\{t\equiv t_0\}$ one constructs the
manifold of orbits, $\Sigma$, of the threading.  Even though $\Sigma$ is
diffeomorphic to the surfaces $\Sigma_t$, $\Sigma$ is not in general given the
same geometry (metric) as any of the $\Sigma_t$.  One gives $\Sigma$ the
threading metric in an attempt to recapture some of the spacetime geometry
associated with the local rest spaces.  Thus, one may think of $\Sigma$ as a
smooth model for the collection of local rest spaces.  $\Sigma$ comes equipped
with the coordinates $x^i$ and the threading metric as a function of not only
the points of $\Sigma$, but also an additional parameter (the parameter along
the original threading curves).  Thus $\Sigma$ has a 1-parameter family of
Riemannian metrics!  This is the beginning of the {\it parametric manifold}
picture of spacetime.

%%%%%%%%%%%%%%%%%%%%%%%%%%%%%%%%%%%%%%%%%%%%%%%%%%%%%%%%%%%%%%%%%%%%%%%%%%%%%%

\Section{Decomposition of Metrics on Fibre Bundles}

The slicing and threading frameworks can be described as part of a single
mathematical structure, a {\it fibre bundle}.  The slicing and threading
metrics, shifts, and lapse functions naturally arise when one examines the
decomposition of a bundle metric in terms of metrics on the base space and the
fibre space.  By choosing the base space and fibre space correctly, one
recovers either the threading or slicing framework.

As the presentations of the slicing and threading viewpoints mainly focused on
the decomposition of the spacetime metric, that will be the main concern in
this section as well.  That is, given a metric on the total space, what
conditions are necessary in order to construct metrics on the base space as
well as the typical fibre.  Since we are interested in this problem in order
to better describe the slicing and threading decompositions, we are really
only interested in local decompositions of the metric.  Thus, we will be
working in a single local trivialisation of the bundle.

Let $\M$ be a fibre bundle with base space $B$, fibre $F$, and continuous,
surjective projection $\pi :\M\to B$.  For the purposes of slicing and
threading one should take $\M$ to be a spacetime and the collection of fibres
$\{\pi^{-1}(x): x\in B\}$ to represent the appropriate foliation (spacelike
hypersurfaces for slicing or timelike curves for threading).  Furthermore, we
have $B=\M / F$ under the equivalence induced by the fibres (i.e.\ for
$p,q\in\M,\ p\sim q \Leftrightarrow \pi(p)=\pi(q)$).  $B$ is called the
{\it manifold of leaves}, and is here the same as the manifold of orbits
introduced earlier.

For the following we will work within a single local trivialisation with
adapted coordinates.  Let $U\subset B$ be a coordinate neighborhood with
coordinates $x^i$ such that $\pi^{-1}(U)\approx U\cross F$.  Furthermore,
within a coordinate neighborhood of $\pi^{-1}(U)$ we may use coordinates
$(x^i, y^{\alpha})$ where $y^{\alpha}$ are coordinates on $F$.

For any $p\in\piinverse(U)$, there exists a natural subspace $V_p\subset
T_p\M$ called the {\it vertical subspace}.  $V_p$ is defined by
$$V_p = \{ X\in T_p\M : \pi\push (X)=0\}.$$
Complementing the notion of vertical, define a subspace $H_p\subset T_p\M$ so
that $T_p\M = V_p \oplus H_p$ and call $H_p$  the {\it horizontal subspace}.
Certainly there are many smooth choices for $H_p$.  If $\M$ has a metric,
$H_p$ may be chosen quite naturally to be the orthogonal complement to $V_p$.

Now, given a metric $g$ on $\M$ is there a natural
choice for metrics $h$ and $k$ on $B$ and $F$ respectively?  Not unless
additional structure on $\M$ is given or we allow for the additional freedom
of a parametric metric.  Let us mention a few of these possibilities in
greater detail.

For $X,\ Y\in T_x B$ there exist unique horizontal lifts of $X$ and $Y$ at
each point $p\in\pi^{-1}(x)$.  Call these lifted vectors $\hat X_p$ and $\hat
Y_p$.  It would seem natural to define $h(X,Y)$ in terms of these lifts.  In
order for $h$ to be well-defined there are various options depending on the
additional structure one is willing to assume.
\item{1.} If $g$ were constant along each fibre, then $g(\hat X_p, \hat Y_p) =
g(\hat X_q, \hat Y_q)$ for all $p,q\in\pi^{-1}(x).$  One could then define
$$h(X,Y) = g(\hat X_p, \hat Y_p)$$ for any $p\in\pi^{-1}(x)$.  Actually, we
can loosen this restriction somewhat.  We only need that $g$ restricted to the
horizontal subspace $H_p$ is constant along each fibre.  This is essentially
Reinhart's condition that $g$ be {\it bundle-like} with respect to the given
foliation
\Ref{{B. Reinhart}, {Ann.\ of Math.}, {\bf 69}, {119--131} (1959).}%
.{}

\item{2.} If there were some preferred global section  $\sigma : B\to\M$
({\it e.g.}\ $F$ is a vector space) one could define
$$h(X,Y)=g(\hat X_{\sigma (x)}, \hat Y_{\sigma (x)}).$$
For instance, $\sigma$ could refer to some initial hypersurface in an initial
value formulation.

\item{3.} One could allow the metric on $B$ to carry extra parameters,
namely $y^{\alpha}$, and define
$$h(X,Y)\at{y^{\alpha}} = g(\hat X_{(x^i,y^\alpha)}, \hat Y_{(x^i,y^\alpha)})$$
and, hence, begin to consider $B$ as a {\it parametric manifold}.

If $h$ has been defined in one of the above situations, the component functions
$h_{ij}$ can be defined and computed.  Suppose the horizontal direction is
defined by the basis
$$\eqalign{H_i &= \dxi +\Gamma^{\alpha}_i \dyalpha .\cr}$$
We denote the {\it horizontal lift} of $\dxi$ by $H_i$, so that $H_i=\hat\dxi$.

One may now define the components of $h$ by
$$\eqalign{\hij &= h(\dxi,\dxj)\cr
	&=g(\hat\dxi,\hat\dxj)\cr
	&=g(\dxi +\Gamma^{\alpha}_i \dyalpha, \dxj +\Gamma^{\beta}_i
\dybeta)\cr
	&=\gij + 2\Gamma^{\alpha}_i g_{j\alpha} +\Gamma^{\alpha}_i
\Gamma^{\beta}_j g_{\alpha\beta}\cr}.$$
We are assuming $h(\dxi,\dxj)$ is well defined, but the functions $\hij$ may
be functions of $y^{\alpha}$ as well as $x^i$ (as in 3).

There are similar obstructions to defining a metric $k$ on $F$.  Vectors
$Z,W\in T_yF$ can be naturally identified with vertical vectors (tangent to the
fibres) of $\M$.  We have many natural embeddings of $F$ into $\M$.  The
problem is which fibre?   As with $h$, we need some additional structure
which allows us a definition in the following sense:
$$k(Z,W)=g(\hat Z, \hat W)$$
where $\hat Z$ and $\hat W$ represent some mapping of $Z$ and $W$ into
vertical vectors of $\M$.  That is, we define $k$ to be a pullback of $g$.
Since $k$ depends on which imbedding of $F$ we use, we can think of $k$ as
being parameterized by the coordinates $x^i$ of $B$.  We therefore define
the components of $k$ by $k_{\alpha\beta}\at{x^i} =
g_{\alpha\beta}\at{x^i}$. That is, one can use the local trivialisations to
pull back the metric $g$ to
a parametric metric on $F$.

We can now write the original metric $g$ of $\M$ in terms of $h$ and $k$.  We
have:
$$(g_{ab}) = \left(\matrix{k_{\alpha\beta} & g_{\alpha j}\cr
\noalign{\hbox{\strut}}
	g_{i \beta} & \hij - 2\Gamma^{\alpha}_i g_{\alpha j} -
\Gamma^{\alpha}_i \Gamma^\beta_j k_{\alpha\beta}\cr}\right)$$
where
$$\eqalign{g_{\alpha j} &= g\left(\dxj,\dyalpha\right)\cr
	&=g\left(\dxj + \Gamma^\beta_j\dybeta , \dyalpha\right) - \Gamma^\beta_j
g_{\alpha\beta}\cr
	&= g\left(H_j, \dyalpha\right) - \Gamma^\beta_j g_{\alpha\beta}\cr.}$$

Since $\M$ has a metric, we may choose our notion of horizontal so that $H_i$
is orthogonal to $\dyalpha$, in which case we have $g(H_j,\dyalpha)=0$ and
$g_{\alpha j} = -\Gamma^\beta_j g_{\alpha\beta} = -\Gamma^\beta_j
k_{\alpha\beta}$.

In this special case $g$ takes the following form:
$$(g_{ab}) = \left(\matrix{k_{\alpha\beta} & -\Gamma^\beta _j
k_{\alpha\beta}\cr
\noalign{\hbox{\strut}}
	-\Gamma^\alpha_i k_{\alpha\beta} & \hij + \Gamma^\alpha_i
\Gamma^\beta_j g_{\alpha\beta}}\right).$$

\proclaim Example \itemno. Threading. \par
{\narrower\narrower
Take $F=\BR$ with coordinate $y^0 = t$ and $B=\Sigma$ to be a 3-dimensional
manifold with coordinates $x^i$ ($i=1,2,3$) as before.  In terms of the above
decompositions, the spacetime metric is of the following form
$$(g_{ab})=\left(\matrix{k_{00} & -\Gamma_j k_{00}\cr
\strut\cr
	-\Gamma_i k_{00} & \hij + \Gamma_i \Gamma_j k_{00}\cr}\right)$$
Since $k_{00}$ represents the squared norm of $\dt$, according to previous
notation $k_{00} = -M^2$.  This decomposition is then precisely the same as
the threading decomposition with $\Gamma_i = M_i$ and where $\hij$ is
the threading metric on the manifold of orbits $\Sigma$.  Thus, the notion of
horizontal is given by $H_i = \dxi + M_i\dt$ which corresponds to the
orthogonal subspace to $\dt$.
\smallskip}

\proclaim Example \itemno. Slicing \par
{\narrower\narrower
By switching the roles of $F$ and $B$ in the above example, one has the
original slicing story.  Let $F=\Sigma$ with coordinates $y^{\alpha} = X^i$
($i=1,2,3$) and $B = \BR$ with a single coordinate $x^0 = t$.  One has:
$$(g_{ab}) = \left(\matrix{k_{ij} & -\Gamma^j k_{ij}\cr
\mathstrut\cr
	-\Gamma^i k_{ij} & h_{00} + \Gamma^i \Gamma^j \kij}\right)$$
As before, $\kij$ is the slicing metric, $\Gamma^i = -N^i$, and $h_{00} =
-N^2$.  Here the horizontal subspace is given by $H_0 = \dt + N^i \dxi$ which
is orthogonal to the hypersurfaces $\Sigma_t$.
\smallskip}

Thus, both the slicing and threading viewpoints are special cases of a more
general decomposition of metrics on fibre bundles.  Furthermore, they are
naturally dual to each other in the sense that the fibre and base space are
interchanged.

%%%%%%%%%%%%%%%%%%%%%%%%%%%%%%%%%%%%%%%%%%%%%%%%%%%%%%%%%%%%%%%%%%%%%%%%%%%%%%

\Section{Parametric Manifolds}

Anyone familiar with the Kaluza-Klein theories of spacetime will notice a
similarity between the threading (1+3) formalism and the standard Kaluza-Klein
(1+4) framework.  Kaluza attempted to describe ordinary 4-dimensional Einstein
gravity and Maxwell electromagnetism by working in a 5-dimensional space
\Ref{{T. Kaluza}, {\it On the Unity Problem of Physics}, reprinted in
{\bf Introduction to Modern Kaluza-Klein Theories}, edited by T. Applequist,
A. Chodos, and P.G.O. Freund, Addison-Welsey, Menlo Park, 1987.}\label\KALUZA
.  Gravity and electromagnetism were then obtained by imposing a
``cylindrical'' condition on the fifth dimension.
The resulting (1+4)-decomposition of a 5-dimensional space is reminiscent of
the threading viewpoint.  In place of the threading metric, Kaluza has the
ordinary Einstein metric of spacetime, and taking the place of the threading
shift is the electromagnetic vector potential.

In \EIN, Einstein and Bergmann reformulated Kaluza's ideas and proceeded to
replace the ``cylindrical'' condition imposed by Kaluza by a ``periodic''
assumption.  In their resulting ansatz lies the beginning of a true parametric
picture of spacetime (although still nestled in the comforts of a
5-dimensional space).  The tensor analysis they described, translated into a
1+3 setting, turns out to be the same as the theory of {\it parametric
manifolds} as presented recently by Perj\'es \PERJES, based on earlier work by
Zel'manov
\Ref{{A. Zel'manov}, {Soviet Physics Doklady} {\bf 1}, 227--230
(1956)}\label\ZELMANOV
.{}

We recently showed
\Ref{Stuart Boersma and Tevian Dray,
{\it Parametric Manifolds I: Extrinsic Approach},
gr-qc/9407011,
J. Math.\ Phys.\ (submitted).} \label\PaperI
how to put this tensor analysis on a rigorous mathematical footing by
generalizing the standard Gauss-Codazzi formalism for projections into an
orthogonal hypersurface.  We now present a summary of those results, stated
explicitly in a threading framework in 1+3 dimensions.  For further details,
see \PaperI.

%%%%%%%%%%%%%%%%%%%%%%%%%%%%%%%%%%%%%%%%%%%%%%%%%%%%%%%%%%%%%%%%%%%%%%%%%%%%%%

Given a timelike vector field $A$ with norm $1/M$, {\it i.e.}\
$\left<A,A\right>=-1/M^2$, introduce a parameter $t$ along the integral curves
of $A$ such that $A = {1\over M^2}\partialt$.  Now introduce (local)
coordinates $x^i$ on the surfaces $\{t=\hbox{constant}\}$.  This is precisely
the threading decomposition of spacetime, adapted to the given family of
timelike curves tangent to $A$ together with a particular choice of
``parameter'' $t$.  In particular, the metric takes the form \threadmet, where
$M$ is to be identified with the threading lapse function, and the $M_i$ are
the components of the threading shift 1-form.

The functions $\hij = \gij + M^2 M_i M_j$ are the components of the threading
metric, and can also be thought of as the nonzero components of the tensor
$$h_{\alpha\beta} = g_{\alpha\beta} + M^2 A_\alpha A_\beta$$
which is associated with the projection operator
$$\proj\alpha\beta = h_\alpha^{\ \beta} =
	\delta_{\alpha}^{\ \beta} + M^2 A_\alpha A^\beta .$$
Any tensor can be projected orthogonally to $A$ by contracting all indices
with $P$; we will denote this operation by the superscript $\perp$.

A {\it parametric manifold} consists of a generic $\{t=\hbox{constant}\}$
hypersurface --- really the manifold of orbits of $A$ --- together with
1-parameter families of ``time-dependent'' projected tensors.  All of the
operations below are covariant in that they take projected tensors to
projected tensors.

In what follows, we will assume that $X$, $Y$, and $Z$ are orthogonal to $A$,
so that {\it e.g.}\ $X^\perp=X$.  We can define the {\it projected covariant
derivative operator} $D$ by
$$D_XY = (\nabla_XY)^\bot .\Eqno$$\label\InducedDeriv
The spatial components of $D_XY$ turn out to be \PaperI
$$(\d XY)^i = X^j \left( Y^i\starry j + \Gamperpu ikj Y^k \right)$$
where we have introduced Perj\'es' {\it starry derivative} notation, namely
$$f\starry i = \partiali f + M_i \partialt f$$
and where we have defined the symbol $\Gamperpu\alpha\nu\beta$ by
$$\Gamperpu\alpha\nu\beta = \proj\gamma\alpha\proj\beta\delta\proj\nu\mu
				\Gamu\gamma\mu\delta. $$

We define the torsion $\tordperp$ of $D$ to be the projection of the
torsion $T$ of $\nabla$.  It turns out that this yields the familiar
coordinate expression \PaperI
$$\tordperp{}^k{}_{ij} = \Gamperpu kij - \Gamperpu kji$$
so that $D$ is torsion free if $\nabla$ is.
However, the coordinate-free version of this equation does not quite take its
usual form, and it is worth considering this carefully.  We have \PaperI
$$\tordperp (X,Y) = \d XY - \d YX - \bracket XY^\perp.$$
Note that only the projected bracket appears here.  If $A$ is
hypersurface-orthogonal, the bracket operation closes --- this is just
Frobenius' theorem.  It can now be shown that if $\nabla$ is both torsion free
and metric compatible, then $D$ is the unique torsion free, metric compatible
(projected) connection associated with the (parametric) metric $h$
\Ref{Stuart Boersma and Tevian Dray,
{\it Parametric Manifolds II: Intrinsic Approach},
gr-qc/9407012,
J. Math.\ Phys.\ (submitted).}\label\PaperII
.  In this case, the connection symbols take the familiar form \PERJES
$$\Gamperpu ijk = \frac 12 h^{im}(h_{mj*k} + h_{mk*j} - h_{jk*m}).
  \Eqno$$\label\GamPerp

However, $A$ will not in general be hypersurface-orthogonal,
and there will also be a parallel component.  We are thus led to define the
{\it deficiency} of $D$ as \PaperI
$$\D(X,Y) = \bracket XY^\top = \bracket XY - \bracket XY^\perp .$$
In components we have
$$\left( \bracket XY^\perp \right)^i = X^m Y^i\starry m - Y^m X^i\starry m$$
and $\D(X,Y)=\D_{ij}X^iY^j \partialt$ with $\D_{ij}=M_{j*i} - M_{i*j}$.
Note that the deficiency does not necessarily vanish even when $\nabla$ is
torsion free.

There are now two natural candidates for the curvature tensor associated with
$D$ \PaperI.  One of these is the {\it Zel'manov} curvature used by Zel'manov
\ZELMANOV\ and Perj\'es \PERJES, as well as (in a 5-dimensional setting) by
Einstein and Bergmann \EIN, namely
$$\left[\sdel k \sdel j - \sdel j \sdel k +\left(M_{j*k} - M_{k*j}\right)
	\dt\right]X_i
		= \zelmanovu rijk X_r\Eqno$$\label\Zel
with components
$$\zelmanovu lkij = \Gamperpu lkj\starry i - \Gamperpu lki\starry j +
\Gamperpu lni \Gamperpu njk - \Gamperpu lnj \Gamperpu nik .$$
Note that the deficiency appears here in much the same way that torsion would.
The other candidate, derived by considering Gauss' equation \PaperI, is
$$\eqalign{\rperpu lkij
	&= \Gamperpu lkj\starry i - \Gamperpu lki\starry j + \Gamperpu lni
		\Gamperpu nkj - \Gamperpu lnj \Gamperpu nki\cr
	&~~~~~+\left(M_{j*i} - M_{i*j}\right) h^{lm}
		\left(M^2 M_{m*k} - M^2 M_{k*m} + \partialt\,h_{km}\right)\cr}
  \Eqno.$$\label\RPerpComp
This expression contains additional terms involving the threading lapse $M$,
but
possesses all the expected symmetries,
\Footnote{The symmetries of this curvature tensor involve the deficiency in
precisely the same way torsion would enter \PaperI.}
which the Zel'manov curvature does not; both contain terms involving the
deficiency.  For a more detailed comparison of these two curvature tensors,
see \PaperI.

%%%%%%%%%%%%%%%%%%%%%%%%%%%%%%%%%%%%%%%%%%%%%%%%%%%%%%%%%%%%%%%%%%%%%%%%%%%%%%

\Section{Discussion}

The theory of parametric manifolds outlined above can also be expressed
intrinsically \PaperII.  The basic idea, as noted by Perj\'es, is to
introduce a {\it parametric structure} on the manifold of orbits, which is
essentially the threading shift 1-form.  This leads to natural generalizations
of Lie differentiation, exterior differentiation, and covariant
differentiation, all based on a nonstandard action of vector fields on
functions.  For further details, see
\PaperII.

In the special case where the threading curves are integral curves of a
Killing vector field, no physical fields depend on the parameter.  In
particular, the projected metric tensor $h$ on the manifold of orbits $\Sigma$
is now an ordinary metric tensor, and it follows immediately from
\GamPerp\ that the projected connection is precisely the Levi-Civita
connection of $h$.  Parametric manifolds thus generalize the formalism of
Geroch
\Ref{R.~Geroch, {\it A Method for Generating Solutions of Einstein's
Equations}, J. Math.\ Phys.\ {\bf 12}, 918 (1971).}
for spacetimes with (not necessarily hypersurface-orthogonal) Killing vectors.

Even in the non-Killing case, if the threading shift 1-form $\omega=M_i\,dx^i$
is independent of the parameter then ``starry'' partial differentiation defines
a connection on the fibre bundle consisting of spacetime over the manifold of
orbits.  Parametric manifolds generalize this to something which is ``almost''
a connection on a fibre bundle.
\Footnote{The missing property is invariance under the action of the group,
which requires the notion of horizontal to be parameter independent.}
As hinted by Perj\'es \PERJES, it may thus be possible to use parametric
manifolds to construct a generalized Yang-Mills theory in which exact symmetry
under the gauge group is not required.

Since they correspond to what a given family of observers actually sees,
parametric manifolds may provide a natural setting for initial-value problems.
For instance, when considering the scalar field on a fixed spacetime
background, formulating the initial-value problem requires one to decompose
the spacetime Laplacian.  Preliminary calculations indicate that, in the
general, non-hypersurface-orthogonal setting, this decomposition is simplest
using the parametric manifold approach
\Ref{Stuart Boersma and Tevian Dray,
{\it Parametric Manifolds and the Scalar Field},
(in preparation).}
.  The standard approach to Killing observers in the exterior of a Kerr black
hole is to use the surfaces of constant time to {\it slice} the spacetime.
This describes physics as seen by the ZAMOs (Zero Angular Momentum Observers)
orthogonal to these surfaces.  The parametric framework would describe instead
the physics actually seen by the Killing observers, since a parametric
manifold captures precisely the geometry of the instantaneous rest spaces of
the given family of observers.  Which approach to use depends on which
questions one is asking.

For the initial value problem in general relativity itself, the situation is
perhaps even more interesting.  The usual constraints relating the extrinsic
curvature and the 3-metric in the slicing approach are derived from the
Gauss-Codazzi equations.  As pointed out above, only one of the candidate
parametric curvature tensors satisfies Gauss' equation.  This may be strong
evidence in support of using this curvature tensor, and may lead to an initial
value formulation of general relativity for appropriate data on the manifold
of orbits of a given vector field.  Work on this issue is continuing.

\bigskip\leftline{\bf ACKNOWLEDGEMENTS}\nobreak

It is a pleasure to thank Bob Jantzen and Zoltan Perj\'es for their
encouragement throughout this project, which builds on earlier results of
theirs.  This work forms part of a dissertation submitted to Oregon State
University (by SB) in partial fulfillment of the requirements for the Ph.D.\
degree in mathematics, and was partially funded by NSF grant PHY-9208494.

%o\vfill\eject

\References

%\Figures

\bye